\documentstyle[amssymb,preprint,aps]{revtex}

\begin{document}
\title{Instantons and the singlet-coupling in the chiral quark model}
\author{T. P. Cheng$^{\ast }$ and Ling-Fong Li$^{\ddagger }$}
\address{$^{\ast }$Department of Physics and Astronomy, University of Missouri, St.\\
Louis, Missouri 63121\\
$^{\ddagger }$Department of Physics, Carnegie Mellon University, Pittsburgh,%
\\
Pennsylvania 15213}
\maketitle

\begin{abstract}
Chiral quark model with a broken U(3) flavor symmetry can be interpreted as
the effective theory of the instanton-dominated non-perturbative QCD. This
naturally suggests the possibility of a negative singlet/octet coupling
ratio, which has been found, in a previous publication, to be compatible
with the phenomenological description of the nucleon spin-flavor structure.
\end{abstract}

\section{Introduction}

When viewed from the perspective of perturbative QCD, some of the
observational data on the nucleon spin/flavor structures appear\ to be
puzzling. The possibility that these non-trivial structures\ originate from
non-perturbative QCD presents itself. Non-perturbative QCD naturally enters
into the study of the nucleon structure because in the hadronic interior $%
\Lambda _{QCD}^{2}\lesssim Q^{2}\lesssim 1\,GeV^{2}$, the QCD gauge coupling
is expected to be large. Features in this intermediate energy regime
correspond to the ``initial distributions'', from which the observed
structures at higher $Q^{2}$'s are related through the standard perturbative
QCD evolutions.

It has been suggested that constituent quarks and internal Goldstone bosons
could be the effective degrees of freedom (DOF) for a simple description of
the phenomena in this non-perturbative region\cite{CL-spin98}. Indeed,
calculations at the level of the non-relativistic chiral quark model $\left(
\chi QM\right) $ have been seen to yield a reasonable account of both the
spin and the flavor structure\cite{EHQ,CL95,CL98-prd}. What is the
theoretical basis for such an effective DOF description? There are several
distinctive theoretical approaches all leading to similar effective
descriptions. They can be differentiated by theoretical (self-consistency)
considerations and by model details.

The suggestion that the instanton configurations dominate the
non-perturbative physics is a particularly attractive possibility\cite
{Diak,Schur,Koch}. It yields the most detailed mechanism for a dynamic
breaking of chiral symmetry, and an effective theory closely resembles the $%
\chi QM$. In this note we shall show that one aspect of the chiral quark
description can be accounted for very naturally by the instanton approach:%
{\em \ i.e}., the phenomenological suggestion of the singlet chiral
meson-quark coupling having an opposite sign from that of the octet
meson-quark coupling \cite{CL95}.

In Sec. II, we shall recall the motivation of working with a $\chi QM$
having a nonet of pseudoscalar mesons, with a negative singlet-coupling as
suggested by phenomenology. In Sec. III, we shall briefly recount the
instanton liquid model; how the 't Hooft determinantal interaction naturally
suggests a negative singlet to octet coupling ratio.

\section{Chiral quark model with a broken U(3) symmetry}

The chiral quark idea\cite{MG-cqm} is that the QCD coupling, as it increases
when proceeding to longer distance scales, could trigger the
non-perturbative phenomenon of spontaneous chiral symmetry breaking before
reaching the confinement radius. Thus, in the intermediate range of $\Lambda
_{QCD}^{2}\lesssim Q^{2}\lesssim \Lambda _{\chi sb}^{2}\approx 1\,GeV^{2},$
the effective DOF are massive constituent quarks and internal Goldstone
bosons (IGBs). For a better understanding, it is important to separate out
the coupling {\em vs} mass effects. For example, in the study of the strange
quark content of the nucleon, the SU(3) symmetric pion-nucleon sigma term
calculation\cite{Cheng76,sigma91} implies a rather large strange quark
content, $\bar{s}>\bar{u},\bar{d},$ while another deduction from the
neutrino charm production (without invoking the $m_{s}=m_{u}=m_{d}$
approximation) suggests \cite{CCFR,str-fun-fits,CL98-prd} a more moderate
situation $\bar{s}\simeq \left( \bar{u}+\bar{d}\right) /2.$

In a previous publication\cite{CL95}, two of us have suggested the
consideration of a chiral quark model with nonet of pseudscalar mesons
(instead of just the usual octet). This was mainly motivated by the
theoretical consideration that in any description of the strong interaction
involving three light flavors of quarks, we would start out with nine
(unmixed) degenerate pseudoscalar mesons --- hence an $U\left( 3\right) $
flavor symmetry. This is the case in the leading $1/N_{c}$ (planar)
approximation ($N_{c}$ being the number of colors). In this limit, the quark
couplings to the singlet meson and to the octet mesons must be equal, $%
f_{1}= $ $f_{8}.\;$This zeroth order approximation misses some essential
physical features: there in no axial anomaly, {\em i.e}. an unbroken axial
U(1) symmetry, and the quark sea is flavor-symmetric, $\bar{u}=\bar{d}=\bar{s%
}$\cite{EHQ,CL95}. Thus any realistic description must involve a broken U(3)
symmetry (due to the higher order non-planar contributions). If one still
wants to work in the simple SU(3) limit\ --- so as to separate out the mass $%
m_{s}>m_{u,d}$ effect (from that of the coupling), it has been suggested
that we should work with two independent couplings, $f_{1}\neq $ $f_{8}.$
One does get a substantially better description of the experimental data
with such a two-parameter fit\cite{CL95}; furthermore, rather surprisingly,
these two couplings are found (for a better phenomenology) to have opposite
signs{\em :} $f_{1}\simeq -$ $f_{8}.$

Why would the coupling sign make a difference in such a simple quark model
calculation? It enters because we must coherently add amplitudes for the
process with different neutral Goldstone boson$\,$intermediate channels\ $%
\left( GB^{0}\right) :$ 
\begin{equation}
q\longrightarrow GB^{0}+q\longrightarrow \bar{q}^{\prime }+q^{\prime }+q
\end{equation}
because they produce the same final states. Hence the relative signs of the $%
\pi ^{0},\,\eta $ and $\eta ^{\prime }$ couplings give rise to an
interference pattern in the production of the $\bar{q}^{\prime }q^{\prime }$
pairs. The amplitude for a neutral GB emission, $q\rightarrow GB^{0}+q,$ is
given by 
\begin{equation}
\left( f_{8}\frac{\pi ^{0}}{\sqrt{2}}+f_{8}\frac{\eta }{\sqrt{6}}+f_{1}\frac{%
\eta ^{\prime }}{\sqrt{3}}\right) \left( \bar{q}q\right)
\end{equation}
where we have not displayed any of the charged GB (as well as the non-flavor
structure) in the coupling. From the quark contents of $\pi ^{0},\,\eta $
and $\eta ^{\prime }$ we can immediately work out the probability for the
quark pair emission, $u\longrightarrow \left( \bar{q}^{\prime }q^{\prime
}\right) +u,$ by the valence $u$ quark\ to be 
\begin{equation}
\left( u\,\bar{u}\right) :\left( d\bar{d}\right) :\left( s\bar{s}\right)
=\left( \frac{2+\zeta }{3}\right) ^{2}:\left( \frac{1-\zeta }{3}\right)
^{2}:\left( \frac{1-\zeta }{3}\right) ^{2}
\end{equation}
where $\zeta =f_{1}/f_{8}.$ One notes, in particular, for $\zeta =-2$ the
interference pattern is such that $u\,\bar{u}\;$pair is not produced by the
valence $u$ quark, while the emission probabilities of $d\bar{d}$ and $s\bar{%
s}$ pairs are non-vanishing and equal in this SU(3) limit. Namely, the quark
sea production always involves a change of the quark flavor: $u\rightarrow d%
\bar{d}u,\,s\bar{s}u,$ $d\rightarrow u\,\bar{u}d,\,s\bar{s}d,$ and $%
s\rightarrow u\,\bar{u}s,\,d\bar{d}s,$ but $u\nrightarrow u\,\bar{u}u,$ $%
d\nrightarrow d\bar{d}d$ and $s\nrightarrow s\bar{s}s,$ etc. This is the
limiting case. The actual phenomenological fits (mainly from the data
showing $\bar{d}>\bar{u}$) suggest more of a value in the neighborhood of $%
\zeta \simeq -1.$

In the chiral quark model, when one includes the $m_{s}>m_{u,d}$
SU(3)-breakings, the singlet $\eta ^{\prime }$ channel is suppressed by the
mass effect of $M_{\eta ^{\prime }}>M_{\eta ,K}>M_{\pi }$\cite{CL98-prd}.
The final result is not particularly sensitive to the $\eta ^{\prime }$
contribution. Thus, without the consideration at an intermediate stage of
the equal mass approximation, the inclusion of the singlet GB contribution
may not be entirely justified on phenomenological ground. However, the
requirement of a (negative) $f_{1}$ seems to suggest that the underlying
theory might be such that the flavor singlet meson is needed in the coupling
scheme, even though its effect is dampened by the SU(3)-breaking mass
effects. [This situation is analogous to the above-discussed issue of
nucleon strange quark content: it's favored by coupling but suppressed by
its large mass.] The relevant point is that all this should be part of the
clues about the correct non-perturbative theory underlying the effective DOF
description.

\section{The Instanton-induced effective interactions}

Non-perturbative QCD being likely to be rather complicated when expressed
directly in terms of the fundamental DOF of (current) quarks and gluons, it
may well be useful to adopt a two-stage approach. In the first stage one
attempts to identify the effective DOF in terms which the physics
description is simple, intuitive and phenomenologically correct. At the
second stage one then tries to work out the relation between these effective
DOF and QCD quarks and gluons. From such a viewpoint the above discussed $%
\chi QM$ is a first-stage description with (constituent) quarks and internal
Goldstone bosons being the effective DOF.

It turns out there are several distinctive non-perturbative QCD approaches,
all leading (at least at the non-relativistic quark model level) to the $%
\chi QM$ as the effective theory\cite{CL-spin98}. One way to differential
the separate approaches is to find model details that can be checked by
experimental measurement. Here we show that the instanton approach naturally
contains the possibility of a negative singlet chiral quark coupling $%
f_{1}/f_{8}\simeq -2$, at least in the SU(3) limit.

Let us first recall that the instanton configuration induces a determinantal
interaction among the light quarks (the 't Hooft interaction\cite{tHooft}): 
\begin{equation}
{\cal H}_{I}=g\det_{i,j}\left[ \bar{q}_{iR}q_{jL}+h.c.\right]
\label{det-int}
\end{equation}
where the flavor indices $i,j=1,2,3$ and $q_{jL}=%
{\frac12}%
(1-\gamma _{5})q_{j},$ etc. In the instanton approach, the light quarks pick
up masses (dynamic symmetry breaking) when propagating in the background of
instanton fields --- they are to be identifies with the constituent quarks.
On the other hand, there are actually no independent propagating
pseudoscalar DOF (i.e. no GB kinetic energy terms), the IGB are just
short-hands for $q\bar{q}$ loop effects: $\bar{q}q$ pairs ``propagate'' by
leaping among states associated with instantons.

This six-quark interaction in (\ref{det-int}) implies that an instanton
absorbs a left-handed quark of each flavor and emits a right-handed quark of
each flavor, $\bar{u}_{R}u_{L}\bar{d}_{R}d_{L}\bar{s}_{R}s_{L}.\;$This
provides a mechanism for produce a negatively polarized quark sea, and (in
the equal mass limit) a flavor structure of $\bar{s}>\bar{d}>\bar{u}$ in the
proton in qualitative agreement with the observed nucleon spin/flavor
structure. In fact we also see that such an interaction would transform a $q%
\bar{q}$ into quark pairs of different flavors --- this is just the $\zeta
=-2$ case discussed in Sec.II.

It may be worthwhile to workout some detail, to see how such opposite signs
arise from the determinantal interaction. Here we shall follow Hatsuda and
Kunihiro\cite{Hatsuda-pr} and use the mean field approximation --- namely a
composite boson field $\Phi _{ij}$ $=\bar{q}_{i}\left( 1-\gamma _{5}\right)
q_{j}$ can be approximated by its (classical) vacuum expectation value, $%
\Phi _{ii}\rightarrow \left\langle \Phi _{ii}\right\rangle \equiv \phi
_{ii}. $ In this way we obtain a set of four-quark vertices from the
determinantal six-quark interaction as\cite{matrix-id} 
\begin{eqnarray}
&&g\left[ \det \Phi +h.c.\right] \stackrel{MFA}{\longrightarrow }  \nonumber
\\
&&g\left[ Tr\left( \phi \Phi ^{2}\right) -%
{\frac12}%
Tr\phi Tr\left( \Phi ^{2}\right) -Tr\left( \phi \Phi \right) Tr\left( \Phi
\right) +%
{\frac12}%
Tr\phi \left( Tr\Phi \right) ^{2}+h.c.\right]  \label{MFAreduction}
\end{eqnarray}
Since we will be working in the SU(3) limit with $\left\langle \bar{u}%
u\right\rangle =\left\langle \bar{d}d\right\rangle =\left\langle \bar{s}%
s\right\rangle \equiv v$, the expectation value matrix reduces to an
identity matrix multiplied by a constant, $\phi =v{\Bbb I}$ and the above
expression is simplified to 
\begin{equation}
g\det \Phi \stackrel{}{\longrightarrow }\frac{vg}{2}\left[ \left( Tr\Phi
\right) ^{2}-Tr\left( \Phi ^{2}\right) \right] =vg\left( \Phi _{0}^{2}-%
{\frac12}%
\Phi _{3}^{2}-%
{\frac12}%
\Phi _{8}^{2}\right)
\end{equation}
where $\Phi _{0}=\left( \bar{u}u+\bar{d}d+\bar{s}s\right) /\sqrt{3},$ $\Phi
_{3}=\left( \bar{u}u-\bar{d}d\right) /\sqrt{2},$ $\Phi _{8}=\left( \bar{u}u+%
\bar{d}d-2\bar{s}s\right) /\sqrt{6}.$ The above expression holds for the
scalar and the pseudoscalar combinations --- they differ by an overall sign.
To cast this in the form of boson quark couplings, we just replace one of
the boson fields by its quark pairs. In this way we see clearly that single
coupling is twice\cite{su3br-f} the negative of the octet couplings $%
\;f_{1}=vg$ and $f_{8}=-vg/2,$ confirming our expectation.

Because the determinantal interaction in (\ref{det-int}) is symmetric under $%
SU_{L}(3)\times SU_{R}(3)$ but not under $U_{A}\left( 1\right) \;$it will
give a mass to the singlet would-be-Goldstone boson (the $\eta ^{\prime }$
meson), thus solving the axial $U_{A}\left( 1\right) $ problem\cite{tHooft}.
We can picture this solution of the $\eta ^{\prime }$ mass problem in more
physical terms: The opposite signs mean that the determinantal interaction
between a quark and an antiquark is an attraction in the octet channel and a
repulsion in the singlet channel. This attraction binds $\bar{q}q$ so
strongly in the octet channel that the resultant state is massless (as
dictated by the gap equation). On the other hand, the corresponding
repulsion in the singlet channel will reduce the binding energy (from the
usual effective four-quark interaction) so that the resultant singlet
pseudoscalar meson is much less tightly bound and is massive. With the U$%
_{A} $(1) solution viewed this way, the opposite singlet-octet coupling
ratio is seen to be related to the determinantal repulsion in the singlet $q%
\bar{q}$ channel and thus ultimately to the resolution of the $\eta ^{\prime
}$ mass problem.

In summary, we have presented an argument suggesting that, in an instanton
dominated non-perturbative QCD, the singlet meson quark coupling naturally
comes out to be negative, which has been found in a previous publication\cite
{CL95} to be compatible with the phenomenology of the nucleon spin-flavor
structure. This, in turn, lends some support to the idea of an
instanton-dominated non-perturbative origin of the hadron structure. In this
connection, we wish to report that another argument for this result, carried
out in the context of anomalous contribution to the singlet axial vector
constant $g_{A}^{0}$\cite{KochVento}, will be presented in a forthcoming
paper\cite{another-arg}.

\begin{description}
\item[Acknowledgment]  One of us (L.F.L.) acknowledges the support from U.S.
Department of Energy, Grant No. DOE-Er/40682/127.
\end{description}


\begin{references}
\bibitem{matrix-id}  One can easily understand the reduction in Eq.(\ref
{MFAreduction}) by recalling an identity for the 3x3 matrix $M:$%
\[
\det M=M^{3}-\left( trM\right) M^{2}+%
{\frac12}%
\left[ \left( trM\right) ^{2}-\left( trM^{2}\right) \right] M. 
\]
The trace of this identity is just the characteristic equation of a cubic
equation.
\bibitem{su3br-f}  When SU(3) breaking in the vacuum expectation values $%
\left\langle \bar{u}u\right\rangle \neq \left\langle \bar{d}d\right\rangle
\neq \left\langle \bar{s}s\right\rangle $ is taken into account, the
magnitude of the ratio $\left| f_{1}/f_{8}\right| $ may well be reduced
(from 2) for a better phenomenological fit.
\end{references}
\end{document}